# An Efficient Likelihood-Based Modulation Classification Algorithm for MIMO Systems

Mohammad Rida Bahloul[1], Mohd Zuki Yusoff[1], Abdel-Haleem Abdel-Aty[2] and M Naufal M Saad[1]
[1]Centre for Intelligent Signal and Imaging Research (CISIR), Department of Electrical and Electronics Engineering, Universiti Teknologi PETRONAS, Perak, Malaysia
[2]Department of Physics and Computers, Faculty of Science, Al-Azhar University, Assiut, Egypt
mohd.bahloul@hotmail.com, mzuki_yusoff@petronas.com.my, halimaty@gmail.com, naufal_saad@petronas.com.my

*Abstract*—Blind algorithms for multiple-input multiple-output (MIMO) signals interception have recently received considerable attention because of their important applications in modern civil and military communication fields. One key step in the interception process is to blindly recognize the modulation type of the MIMO signals. This can be performed by employing a Modulation Classification (MC) algorithm, which can be feature-based or likelihood-based. To overcome the problems associated with the existing likelihood-based MC algorithms, a new algorithm is developed in this paper. We formulated the MC problem as maximizing a global likelihood function formed by combining the likelihood functions for the estimated transmitted signals, where Minimum Mean Square Error (MMSE) filtering is employed to separate the MIMO channel into several sub-channels. Simulation results showed that the proposed algorithm works well under various operating conditions, and performs close to the performance upper bound with reasonable complexity.

## I. INTRODUCTION

Communications over multiple-input multiple-output (MIMO) wireless channels have been the focus of intense research over the last decade. This is mainly due to the rapid development of the high speed broadband wireless communication technologies. The use of multiple-antennas for both transmission and reception can significantly improve the wireless link performance through the capacity and diversity gains, resulting in much more reliable transmission relative to the conventional single-input single-output (SISO) systems. Because of these advantages, MIMO has become one of the most important parts of the modern wireless communication standards, such as IEEE 802.11n, 802.16e, 3rd Generation Partnership Project–Long Term Evolution (3GPP LTE), IEEE 802.20 and 802.22 [1].

Recently, blind algorithms for MIMO signals interception have received considerable attention of many researchers (see, e.g., [2]–[4], and the references therein). This is due to their important applications in modern civil and military communication fields. One key step in the interception process is to blindly recognize the modulation type of the MIMO signals, which can be performed by employing a Modulation Classification (MC) algorithm [5].

In general, there are two classes of MC algorithms for MIMO systems: Feature Based (FB) (e.g., [6]) and likelihood Based (LB) algorithms (e.g., [7]). FB algorithms use pre-defined set of features in combination with classification systems to achieve the modulation classification. Even though these algorithms are generally easy to implement and relatively robust to model mismatch such as phase offsets and timing errors, they are not optimal in the Bayesian sense and often require an offline phase to train a classification system. LB algorithms, on the other hand, do not require an offline phase and can be optimal in the Bayesian sense, but at the price of some increase in the computational complexity [5]. These algorithms compute the likelihood functions for the received or the recovered signals under different modulation hypotheses, and make classification decisions based on the maximum of these functions. Depending on the way the data and unknown parameters such as the MIMO channel matrix and noise variance are treated when calculating the likelihood functions, LB algorithms for MIMO systems can be generally grouped into two main classes: Average Likelihood Ratio Test (ALRT)-based and Hybrid Likelihood Ratio Test (HLRT)-based algorithms. The former treat the data and unknown parameters as random variables with known Probability Density Functions (PDFs) to be averaged over. Hence, they need prior knowledge about the PDFs of these unknowns, which may not necessarily be practical or available in case of the unknown communication parameters [5]. The HLRT-based algorithms, on the other hand, are more practical since they treat the unknown parameters as unknown deterministics to be estimated and the data only as random variables to be averaged over.

In this paper, we focus on LB MC algorithms for MIMO systems rather than FB ones, which have been extensively investigated in the last few years, since they can provide optimal solutions when the unknown parameters or their reliable estimates are available at the receiver side. Several LB MC algorithms for MIMO systems [7]–[10] have been found in the literature. In [7], two LB MC algorithms for MIMO systems are proposed. The first one is an ALRT-based algorithm developed under the assumption that a perfect knowledge of the MIMO channel matrix is available at the receiver side. The second one, which is more practical, is an HLRT-based algorithm developed to relax the previous assumption; Independent Component Analysis (ICA)-based channel estimation technique is employed to estimate the MIMO channel matrix. These algorithms suffer from very high computational complexity which makes them impossible for real-time operation, especially when high-order modulation schemes are considered and/or the number of the transmitting antenna is large. Instead of utilizing an ICA-based channel estimation technique as other LB algorithms, HLRT-based MC algorithms with an Expectation-Maximization (EM)-based channel estimation technique are developed in [8], [9]. Hence, unlike the MC algorithms with ICA, these algorithms do not impose the requirement that the number of receiving antennas must exceed the number of transmitting antennas to be applicable. However, they tend to be as computationally expensive as the algorithms introduced in [7]. Furthermore, the channel estimation technique employed in this study may fail to accurately capture the MIMO channel, since it is very sensitive to the initialization conditions and may easily get stuck in local optima [11]. In [10], an efficient HLRT-based MC algorithm is proposed. Instead of calculating the likelihood function for the received MIMO signal like other LB algorithms which is highly complex, this algorithm calculates the likelihood functions for the estimated transmitted (i.e., recovered) signals for each possible modulation scheme and combines them to construct a global likelihood function. The classification decision is then made based on the maximum of the overall likelihood function. However, the likelihood functions are combined in this algorithm using the product as the combination rule, hence, its overall performance is dominated by the worst likelihood function value of all recovered signals. If at least one recovered signal suffers from high noise levels, this algorithm suffers from significant performance degradation. Furthermore, this algorithm assumes that the noise at the recovered signals is independent and has the same variance for each signal, which is an unrealistic assumption. The noise at the estimated signals in practice is correlated and has different variances [12].

To overcome the problems associated with the previous likelihood-based MC algorithms for MIMO systems, a new HLRT-based algorithm is introduced in this study. It considers the real wireless communication scenarios where the receiver has no knowledge of the channel state; an ICA-based channel estimation technique is employed to estimate it. The proposed algorithm is highly efficient since the MC problem, as the algorithm in [10], is formulated as maximizing global likelihood functions formed by combining the likelihood functions computed for the estimated transmitted signals. However, the proposed algorithm employs Minimum Mean Square Error (MMSE) filtering after performing the ICA analysis to separate the MIMO channel into several sub-channels; the channel after the MMSE filter can be well treated as an additive white Gaussian noise (AWGN) channel [13], [14]. Furthermore, instead of using the product as the combination rule, a weighted sum rule is proposed in this study to combine the likelihood functions under each hypothesis; the weighting coefficient vector is defined as the set that maximizes the combined likelihood function. Hence, the performance of the algorithm is not dominated by the worst likelihood function value of all estimated signals as in [10], and thus more reliable classification results can be achieved.

The rest of this paper is organized as follows. Section II introduces the signal model and lists the assumptions. Section III describes the proposed MC algorithm in detail. Section IV presents the simulation results, and finally, section V concludes the whole paper.

## II. SIGNAL MODEL

Let us consider a spatial multiplexing MIMO system equipped with $M_T$ transmitting antennas and $M_R \geq M_T$ receiving antennas. Under the assumption that the channel is time invariant and frequency flat, the received symbol vector at time instant $k$ can be expressed as

$$\mathbf{r}(k) = \mathbf{H}\mathbf{s}(k) + \mathbf{n}(k), \qquad (1)$$

where $\mathbf{r}(k)$ is an $(M_R \times 1)$ received symbol vector at time instant $k$ under the assumption of perfect frequency and time synchronization at the receiver side; $\mathbf{s}(k)$ is the $(M_T \times 1)$ transmitted symbol vector at time instant $k$ whose elements are assumed to be independent and identically distributed (i.i.d.); $\mathbf{n}(k)$ is an $(M_R \times 1)$ additive background noise vector at time instant $k$ corresponding to the spatially and temporally white circularly symmetric complex Gaussian noise with zero mean and variance $\sigma_n^2$; and $\mathbf{H}$ is the $(M_R \times M_T)$ MIMO channel matrix whose entries correspond to the complex path gain between the transmitting and receiving antennas.

A Rayleigh fading channel is considered in this study; thus all complex entries of $\mathbf{H}$ are assumed to follow a circularly symmetric complex Gaussian distribution with zero mean and unit variance.

Without loss of generality, it is assumed that the signal transmitted from each antenna has a unity average power. Consequently, the average SNR at the receiver side can be expressed as SNR=$10 \log(M_T/\sigma_n^2)$. It is further assumed that the number of transmitting antennas $M_T$ and the noise variance $\sigma_n^2$ are either perfectly known or correctly estimated at the receiver side. This is a common assumption in the literature [6]–[10]; where these parameters can be obtained using the techniques outlined in [15].

## III. PROPOSED CLASSIFICATION ALGORITHM

To overcome the problems associated with the existing likelihood-based algorithms, a new HLRT-based algorithm for MIMO systems is introduced in this section. The proposed algorithm has two main stages. In the first stage, blind channel estimation and equalization are performed by using ICA followed by MMSE linear filtering; where estimates of the channel matrix and the transmitted signals are obtained at the output of this stage. Then, in the second stage, the likelihood functions are calculated for the recovered signals under different modulation hypotheses using the estimates obtained in the first stage, and combined to construct a global likelihood function. The classification decision is then made based on the maximum of this function. The stages of the algorithm are discussed in detail below.

### A. Blind Channel Estimation and and Equalization

Since the received signal vector components are instantaneous linear mixtures of the transmitted signal vector components plus white noise, blind channel estimation and equalization are required to recover the transmitted signals from the received mixture.

ICA [16], which is used in this study, is the conventional and most widely used approach for solving this problem. It maximizes an objective function that characterizes the statistical independence of the transmitted signals. However, the ICA is followed in this paper by a linear filter based on the MMSE criterion to cope with residual interference.

Several algorithms based on different criteria have been proposed so far to perform ICA, such as Infomax ICA [17] and Second Order Blind Identification (SOBI) [18]. Dues to its fast convergence rate, and satisfactory separation performance in many applications [19], the Joint Approximate Diagonalization of Eigen-matrices (JADE) [20] algorithm is employed in this study to perform ICA.

In practice, ICA enables us to estimate the channel matrix and the transmitted signals blindly up to a permutation and a phase ambiguity. Since the ordering of the signals at the receiver side is not important for MC algorithms, the first ambiguity (i.e., permutation ambiguity) is not a problem and does not affect the performance [7]. However, the phase ambiguity must be estimated and compensated for prior to the calculation of the likelihood functions since it may significantly affect the classification results [7]. Thus, initial estimates of the channel matrix $\widehat{\mathbf{H}}$ and the transmitted signals $\tilde{\mathbf{s}}(k)$ are first obtained with JADE. Then, power-law estimator [21], which is a simple and blind phase estimator, is employed to estimate and compensate the phase ambiguity inherent in the estimated transmitted signals. Under hypothesis $H_A$, the phase offset estimate corresponding to the $m$-th transmitted signal can be express as [22]

$$\hat{\theta}_m^{(A)} = \frac{1}{P} \arg\left(\mu_A^{(P,P)} \sum_{k=1}^{N} (\tilde{s}_m(k))^P\right), \quad (2)$$

where $P = 4$ for Quadrature Amplitude Modulations (QAM) schemes and equals the modulation order for any Phase-Shift-Keying (PSK) schemes, and $\mu_A^{(P,P)}$ is the moment of order $P$ with $P$ conjugates corresponding to the modulation scheme $A$ computed over ideal noise-free channels. The phase corrected channel estimate under hypothesis $H_A$ can be then expressed as

$$\widehat{\mathbf{H}}^{(A)} = \widehat{\mathbf{H}}\, \widehat{\boldsymbol{\theta}}^{(A)}, \quad (3)$$

where $\widehat{\boldsymbol{\theta}}^{(A)}$ is a diagonal matrix whose diagonal elements are the phase estimates for the $M_T$ transmitted signals under hypothesis $H_A$.

In this work, we choose to employ an MMSE filter after performing the ICA analysis and resolving the phase ambiguity to separate the MIMO channel into several sub-channels. The MMSE filter matrix under hypothesis $H_A$ can be given by [23]

$$\mathbf{G}^{(A)} = \widehat{\mathbf{H}}^{(A)}\left((\widehat{\mathbf{H}}^{(A)})^H \widehat{\mathbf{H}}^{(A)} + \sigma_n^2\, \mathbf{I}_{M_T}\right)^{-1}. \quad (4)$$

Accordingly, the estimate symbol vector at time instant $k$ $(1 \leq k \leq N)$ under hypothesis $H_A$ can be expressed as

$$\begin{aligned}\hat{\mathbf{s}}^{(A)}(k) &= (\mathbf{G}^{(A)})^H \mathbf{r}(k) \\ &= (\mathbf{G}^{(A)})^H\, \widehat{\mathbf{H}}^{(A)} \mathbf{s}(k) + (\mathbf{G}^{(A)})^H \mathbf{n}(k) \\ &= (\mathbf{G}^{(A)})^H \widehat{\mathbf{H}}^{(A)} \mathbf{s}(k) + \mathbf{v}^{(A)}(k),\end{aligned} \quad (5)$$

where $\mathbf{v}^{(A)}(k) = \left[v_1^{(A)}(k), \ldots, v_{M_T}^{(A)}(k)\right]^T$ is the filtered noise vector at time instant $k$ under hypothesis $H_A$. Denoting $\hat{\mathbf{h}}_i^{(A)}$ $(1 \leq i \leq M_T)$ and $\mathbf{g}_i^{(A)}$ as the $i^{th}$ column of $\widehat{\mathbf{H}}^{(A)}$ and $\mathbf{G}^{(A)}$, respectively, we can express the symbol estimate for the $i^{th}$ signal at time instant $k$ under hypothesis $H_A$ as

$$\begin{aligned}\hat{s}_i^{(A)}(k) &= (\mathbf{g}_i^{(A)})^H \hat{\mathbf{h}}_i^{(A)} s_i(k) \\ &\quad + \sum_{j=1, j\neq i}^{M_T} (\mathbf{g}_i^{(A)})^H\, \hat{\mathbf{h}}_j^{(A)} s_j(k) + v_i^{(A)}(k) \\ &= (\mathbf{g}_i^{(A)})^H \hat{\mathbf{h}}_i^{(A)} s_i(k) + w_i^{(A)}(k),\end{aligned} \quad (6)$$

where $w_i^{(A)}(k)$ is the distortion for the $i^{th}$ signal under hypothesis $H_A$ which consists of the residual interference and the noise.

As the distribution of $w_i^{(A)}(k)$ can be well approximated by a Gaussian [13], we assume that the terms in $w_i^{(A)}(k)$ make a complex Gaussian distribution with zero mean and variance $\sigma_{w,i}^2$ which can be expressed as [14]

$$\sigma_{w,i}^2 = (\mathbf{g}_i^{(A)})^H\, \hat{\mathbf{h}}_i^{(A)}\left(\mathbf{1} - (\mathbf{g}_i^{(A)})^H \hat{\mathbf{h}}_i^{(A)}\right). \quad (7)$$

### B. Likelihood Function Calculation

Let $H_A$ denote the hypothesis that the modulation scheme $A$ $(A \in \mathcal{A})$ is used to represent the data. By using the HLRT approach and the facts that the transmitted signals are independent from symbol to symbol, we can express the likelihood function for the $i^{th}$ estimated transmitted signal under hypothesis $H_A$ as

$$f\left(\hat{\mathbf{s}}_i^{(A)}\right) = \prod_{k=1}^{N} \int f\left(\hat{s}_i^{(A)}(k)\middle| s_i^{(A)}(k)\right) P\left(s_i^{(A)}(k)\right) ds_i, \quad (8)$$

where $\hat{\mathbf{s}}_i^{(A)} = \left[\hat{s}_i^{(A)}(1), \ldots, \hat{s}_i^{(A)}(N)\right]$ is the $i^{th}$ $(1 \leq i \leq M_T)$ estimated transmitted signal; $s_i^{(A)}(k)$ is an unknown symbol transmitted from antenna $i$ at time instant $k$ belonging to the modulation scheme $A$; $f\left(\hat{s}_i^{(A)}(k)\middle| s_i^{(A)}(k)\right)$ is the likelihood function for the $\hat{s}_i^{(A)}(k)$ conditioned on $s_i^{(A)}(k)$; and $P(s_i^{(A)}(k))$ is the a priori probability of $s_i(k)$ under hypothesis $H_A$.

Without loss of generality, we assume that all constellation points of the modulation scheme have the same a priori probability. Hence, $P(s_i^A(k)) = 1/|A|$; where $|A|$ denotes the number of possible constellation points for modulation scheme $A$ (e.g., $|A| = 16$ in the case of 16-QAM). Using this assumption and (6), we can write (8) as

$$f\left(\hat{\mathbf{s}}_i^{(A)}\right) =$$

$$\frac{1}{(|A|\pi\sigma_{w,i}^2)^N} \prod_{k=1}^{N} \sum_{s_i(k) \in A} \exp\left[-\frac{\left|\hat{s}_i^{(A)}(k) - (\mathbf{g}_i^{(A)})^H \hat{\mathbf{h}}_i^{(A)} s_i(k)\right|^2}{\sigma_{w,i}^2}\right], \quad (9)$$

The likelihood functions computed for the recovered signals are combined in this study under each modulation hypothesis to construct a global likelihood function in a way similar to the algorithm in [10]. However, instead of using the product rule as the combination rule, the weighted sum rule is proposed to be used to combine the likelihood functions under each hypothesis. Hence, the combined likelihood function under hypothesis $H_A$ can be expressed as

$$f\left(\{\hat{\mathbf{s}}_i^{(A)}\}_{i=1}^{M_T}\right) = \sum_{i=1}^{M_T} \beta_i f\left(\hat{\mathbf{s}}_i^{(A)}\right) = \boldsymbol{\beta}^T\, \mathbf{f}^{(A)}, \quad (10)$$

where $\mathbf{f}^{(A)} = \left[ f(\hat{\mathbf{s}}_1^{(A)}), \ldots, f(\hat{\mathbf{s}}_{M_T}^{(A)}) \right]^T$ is the vector of the likelihood functions computed for the estimated transmitted signals under hypothesis $H_A$, and $\boldsymbol{\beta} = \left[ \beta_1, \ldots, \beta_{M_T} \right]^T$ is the vector of the weighting coefficients corresponding to these functions whose norm is bounded by one, i.e., $\|\boldsymbol{\beta}\|_2 = 1$. Note in (10) that small values of $f(\hat{\mathbf{s}}_i^{(A)})$ due to high noise levels at the estimated signals do not highly affect the whole expression, such as when using the product as the combination rule.

The optimal weighting coefficient vector is defined in this study as the set that maximizes $f(\{\hat{\mathbf{s}}_i^{(A)}\}_{i=1}^{M_T})$[1]. From Cauchy-Schwarz inequality, the optimal weighting coefficient corresponding to $f(\hat{\mathbf{s}}_i^{(A)})$ can be obtained as

$$\beta_i^{opt} = \sqrt{\frac{f(\hat{\mathbf{s}}_i^{(A)})}{\sum_{i=1}^{M_T} f(\hat{\mathbf{s}}_i^{(A)})}} \qquad (11)$$

The likelihood function in (10) will be evaluated for all candidate modulation schemes and the one producing the largest result will be chosen as the scheme for the received MIMO signal; that is

$$\hat{A} = \arg \max_{A \in \mathcal{A}} \left\{ f(\{\hat{\mathbf{s}}_i^{(A)}\}_{i=1}^{M_T}) \right\}, \qquad (12)$$

where $\hat{A}$ is the estimate of the modulation. It is important to note that, since the solution of (11) does not directly maximize the probability of correct classification, the proposed algorithm is not necessarily optimal in the Bayes sense. However, this issue is only important if the communication parameters are perfectly known or accurately estimated at the receiver side, which is an unrealistic assumption for real wireless communication scenarios. Under realistic assumptions on the wireless MIMO channel, the proposed HLRT-based algorithm shows a performance close to the performance upper bound while having a significantly lower computational complexity. The performance of the proposed algorithm is presented in the next section.

## IV. RESULTS AND DISCUSSION

Extensive Monte Carlo simulations were conducted in MATLAB® to evaluate the performance of the proposed algorithm. These simulations have been performed for different SNR values ranging from -10 dB to 15 dB. This range is of interest because it is typical for cognitive radio and military communication systems which are the most common application areas for MC.

Four modulation schemes are considered in this study, which are BPSK, QPSK, 8-PSK, and 16-QAM. They are nowadays the most common schemes employed for transmission in modern civilian and military communication systems.

As in most literature on MC for MIMO systems, the probability of correct classification averaged over all the considered modulation schemes ($P_{cc}$) is used as a measure of the performance of the algorithm. Five hundred Monte Carlo trials were performed for each considered modulation scheme and SNR value; where $P_{cc}$ was obtained as the ratio of the number of trials at which the modulation scheme had been correctly classified to the total number of trials. For all Monte Carlo trials, unless otherwise noted, $N=512$ i.i.d symbols per transmit antenna was considered as an observation length and $M_T=2$, and $M_R=4$ as MIMO antenna configuration.

[1] The optimal weighting coefficient vector can be also defined as the set that maximizes the probability of correct classification. However, unfortunately, an analytical expression for this probability could not be found in the case of MC for MIMO systems [24].

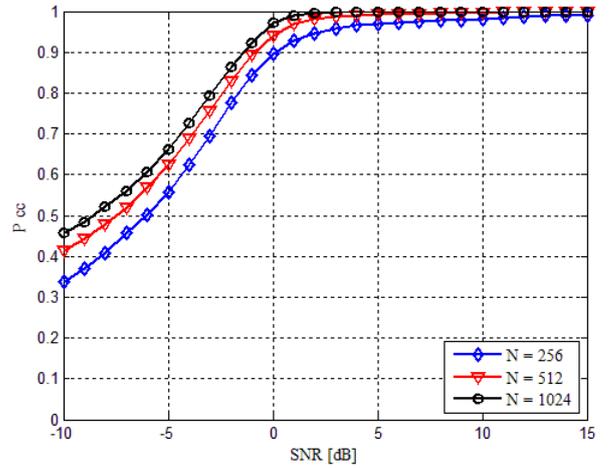

Fig. 1. Average probability of correct classification versus SNR for different observation interval lengths.

Fig. 1 illustrates the effect of the observation length $N$ on the performance of the proposed MC algorithm, where $N$ is set to 256, 512, and 1024, respectively. As shown from these results, the performance improves as the $N$ increases. This is because, as the observation length increases, the accuracy of the channel and phase estimates also increase, resulting in an improvement in the algorithm performance. Note that the proposed algorithm performs well even for very short $N$.

In addition to the observation length, the MIMO antenna configuration is also an important parameter that affects the classification performance of MC algorithms for MIMO systems. Fig. 2 illustrates the effect of the MIMO antenna configurations on the performance of the proposed MC algorithm, where $M_T$ is set to 2, and $M_R$ to 4, 6, and 8, respectively. As shown form these results, the algorithm performance improves as $M_R$ increases (for fixed $M_T$). This is because of the fact that the Gaussian assumption, which is made for the residual interference plus noise at the output of the MMSE filter, is more valid for higher order MIMO configurations [23]. Note that the proposed algorithm performs well for all the investigated configurations.

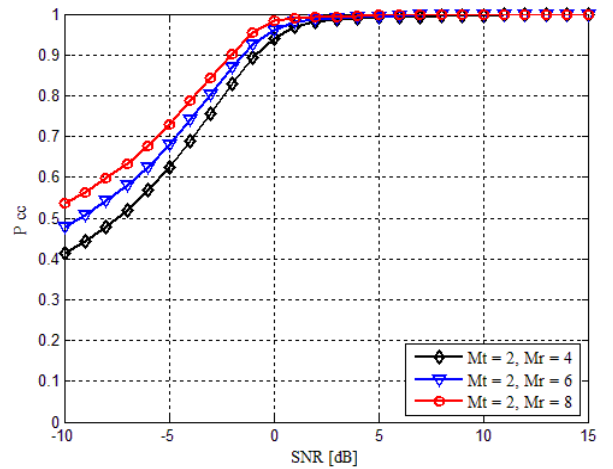

Fig. 2. Average probability of correct classification versus SNR for different MIMO antenna configurations.

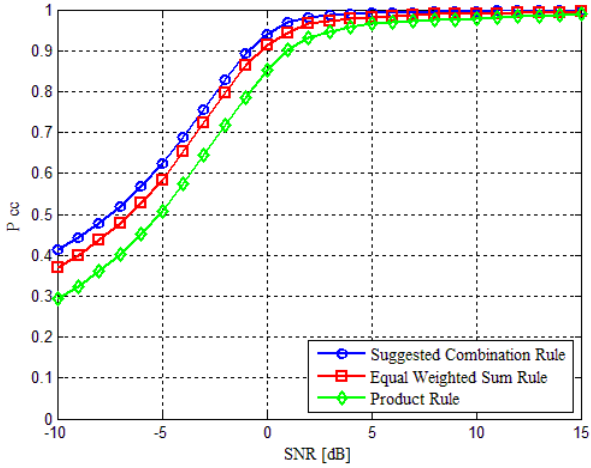

Fig. 3. Average probability of correct classification versus SNR for the combination scenarios under test.

Fig. 3 compares the performance of the proposed algorithm with the suggested combination rule to its performance when using the equal weighted sum rule which is a special case of the proposed rule (i.e., when setting the weighting coefficients $\{\beta_i\}_{i=1}^{M_T}$ in (10) to $1/M_T$) or the product rule (i.e., as the algorithm in [10]) as the combination method. As seen, the performance of the proposed algorithm when employing the proposed combination rule is clearly better than the performance when employing the other rules. This is because the proposed rule assigns weights to the individual probabilities in such a way that the overall probability corresponding to the modulation scheme under test is maximized.

As mentioned in [25], a direct comparison with other studies is very difficult in modulation classification. This is due to a number of reasons, e.g. considering different modulation sets and handling different unknown communication parameters.

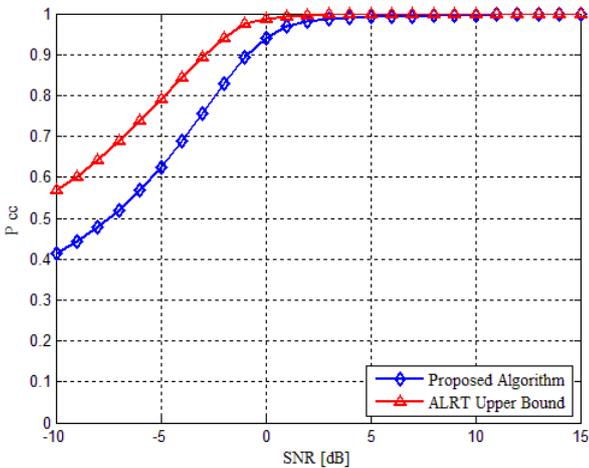

Fig. 4. Performance comparison of the proposed algorithm with that of the ALRT-UB.

However, in the literature, Choqueuse et al. developed an ALRT-based algorithm [7] for MIMO systems under the assumption of perfect knowledge of the communication parameters at the receiver side; thus, it provides an upper performance bound for the MC problem. Hence, this bound, referred to as the ALRT-Upper Bound (ALRT-UB), is used in this paper as a performance benchmark. Fig. 4 compares the performance of the proposed algorithm with that of the ALRT-UB under the same simulation conditions. As expected, ALRT-UB achieves better performance than the proposed algorithm since it assumes a perfect knowledge of the channel matrix and uses the likelihood function of the received signal to perform the classification. Note that the performance difference is not that significant. For instance, at $P_{cc}$ equal to 90%, the performance difference is only about 2 dB. However, the complexity order for the proposed algorithm, which can be approximated by $\mathcal{O}(N|A|M_T)$ given $H_A$, is significantly lower than that for the ALRT-UB, whose complexity order is $\mathcal{O}(N|A|^{M_T})$ given $H_A$ [10]. Accordingly, the proposed algorithm is suitable for the practical and real-time applications.

V. CONCLUSION

In this paper, a low-complexity HLRT-based MC algorithm for MIMO systems is presented. We simplified the formulation of the MC problem by treating the individual estimated transmitted signals as independent processes, where a linear filter based on the MMSE criterion is employed to separate the MIMO channel into several sub-channels. Furthermore, we introduce a weighted sum rule to combine the likelihood functions of the estimated transmitted signals under each hypothesis. It assigns weights to the individual likelihood functions in such a way that the overall function corresponding to the modulation scheme under examination is maximized. The proposed MC algorithm shows good performance with reasonable complexity under various operating conditions, and its performance is comparable to the performance upper bound.